# Robust Semi-Supervised CT Radiomics for Lung Cancer Prognosis: Cost-Effective Learning with Limited Labels and SHAP Interpretation

Mohammad R. Salmanpour [1,2,3*], Amir Hossein Pouria [4], Sonia Falahati [3,5], Shahram Taeb [6], Somayeh Sadat Mehrnia[7], Mehdi Maghsudi[3], Ali Fathi Jouzdani [8], Mehrdad Oveisi [3,9], Ilker Hacihaliloglu [1,10], Arman Rahmim [1,2,11]

[1]Department of Radiology, University of British Columbia, Vancouver, BC, Canada
[2]Department of Integrative Oncology, BC Cancer Research Institute, Vancouver, BC, Canada
[3]Technological Virtual Collaboration (TECVICO Corp.), Vancouver, BC, Canada
[4]Department of Computer Engineering, Amirkabir University of Technology, Tehran, Iran
[5]Electrical and Computer Engineering Department, Nooshirvani University of Technology, Babol, Iran
[6]Department of Radiology, School of Paramedical Sciences, Guilan University of Medical Sciences, Rasht, Iran
[7]Department of Integrative Oncology, Breast Cancer Research Center, Motamed Cancer Institute, ACECR, Tehran, Iran
[8] NAIRG, Department of Neuroscience, Hamadan University of Medical Sciences, Hamadan, Iran
[9]Department of Computer Science, The University of British Columbia, Vancouver, BC, Canada
[10]Department of Medicine, University of British Columbia, Vancouver, BC, Canada
[11]Department of Physics & Astronomy, University of British Columbia, Vancouver, BC, Canada

(*) Corresponding Author: Mohammad R. Salmanpour, PhD; msalman@bccrc.ca

## Abstract

**Background:** CT imaging is essential for lung cancer management, while offering detailed visualization and valuable information for AI-guided prognosis. Meanwhile, supervised learning (SL) models require extensive labeled data, limiting their real-world utility where annotations are scarce.
**Methods:** We analyzed CT scans from 977 patients across 12 public/private datasets, extracting 1,218 radiomics features using Laplacian of Gaussian and wavelet filters via standardized PyRadiomics. Dimensionality was reduced using 56 feature selection and attribute extraction algorithms, and 27 classifiers were benchmarked. Semi-supervised learning (SSL) framework with pseudo-labeling utilized 478 unlabeled and 499 labeled cases. Model sensitivity was assessed in three scenarios: varying labeled data in SL, increasing unlabeled data in SSL, and jointly scaling both from 10% to 100%. SHAP analysis explained and interpreted the top model predictions. Five-fold cross-validation and external testing (in two cohorts) were performed.
**Results:** SSL outperformed SL across all metrics, improving overall survival prediction by up to 17%. The top SSL model (Feature Importance by Random Forest+XGBoost) achieved 0.90±0.01 accuracy in cross-validation and 0.88±0.01 externally. SHAP revealed enhanced feature discriminability in SSL and SL, particularly for class 1 (survival>4 years), and helped explain model decisions. SSL maintained strong performance even with only 10% labeled data. Both SSL-based scenarios demonstrated more stable performance compared to SL, with lower variance across external testing, emphasizing SSL's robustness and cost-efficiency.
**Conclusion:** We introduced a cost-effective, stable, and interpretable SSL framework for CT-based survival prediction in lung cancer, enhancing performance, generalizability, and clinical readiness by integrating SHAP explainability and leveraging unlabeled data.

## 1. Introduction

Lung cancer (LCa) remains the leading cause of cancer-related deaths globally, with Non-Small Cell Lung Cancer (NSCLC) accounting for approximately 85% of all cases; despite recent therapeutic advancements, the five-year survival rate for metastatic NSCLC remains dismal—estimated at only 3.2% [1]. Accurate prediction of overall survival (OS) is therefore crucial to inform risk-adaptive, personalized treatment strategies. However, prognostic modeling in NSCLC is challenging due to the complex heterogeneity of tumor biology and imaging patterns [2].

Computed tomography (CT) plays a central role in lung cancer (LCa) screening and staging due to its accessibility, speed, and resolution. Compared to positron emission tomography/computed tomography (PET/CT), which is often used for metabolic assessment, CT is more widely available and cost-effective. In North America and globally, PET/CT suffers from higher patient doses [3], high diagnostic costs [4], limited availability and long waiting times [5]—especially in community hospitals and resource-limited settings—highlighting the urgent need for effective CT-only prediction tools. Therefore, enhancing the prognostic utility of CT imaging could enable broader implementation of AI-driven precision medicine in LCa care [6].

Recent advances in artificial intelligence (AI), particularly radiomics and machine learning (ML), have shown significant potential for extracting high-dimensional quantitative features from medical images, offering deeper insight into tumor phenotype and prognosis [7, 8, 9]. Handcrafted radiomics features (RF) and deep RFs have been used to characterize tumor heterogeneity and predict patient outcomes [10, 11]. However, most radiomics-based studies rely on fully annotated datasets, which are often scarce, expensive, and time-consuming to generate in real-world clinical settings [12]. This limitation affects the generalizability and scalability of traditional supervised learning (SL) models.

Semi-supervised learning (SSL) presents a promising alternative by leveraging both labeled and unlabeled data, thus mitigating the bottleneck of annotation scarcity [13]. SSL frameworks, such as those incorporating pseudo-labeling, have demonstrated superior performance over SL models in various clinical tasks, including cancer detection and prognosis [14]. Prior work has explored the fusion of RFs and deep RFs for OS prediction in PET/CT images using SSL strategies [15]. Nevertheless, most existing SSL applications are limited by small sample sizes, modality dependency (e.g., PET/CT), or lack of external validation, which hinders reproducibility and clinical translation.

This study introduces a scalable, multicenter, CT-only SSL framework for robust OS prediction in NSCLC patients. Unlike prior studies dependent on multimodal imaging, our approach leverages only CT scans and a diverse pool of labeled and unlabeled data collected from 12 public and private datasets. By integrating a comprehensive radiomics pipeline with 56 dimensionality reduction methods (FSAs/AEAs) and 27 classifiers, we evaluate model performance under both SL and SSL frameworks. The framework is validated using rigorous 5-fold cross-validation and two external test cohorts, ensuring generalizability across patient populations. This work addresses key gaps in the field by demonstrating that SSL can maintain high predictive performance with limited labeled data, reducing dependence on costly annotations and enabling broader deployment in real-world screening environments. Our results emphasize the potential of CT-based SSL models in advancing scalable and equitable AI applications in LCa prognosis.

## 2. Materials and Methods

### 2.1. Patient Data

Clinical data, CT images, and manually delineated lesion masks (elaborated in section 2.2 (i and ii)) were collected from a total of 977 patients across 12 publicly and privately available datasets, all sourced from The Cancer Imaging Archive (TCIA). The datasets included: LCTSC (35 out of 60 patients) [16], LIDC-IDRI (279 out of 1010 patients) [17], Lung-Fused-CT-Pathology (6 patients) [18], LungCT-Diagnosis (48 out of 61 patients) [19], NSCLC-Radiogenomics (34 out of 211 patients) [20], NSCLC-Radiomics (417 out of 422 patients) [21], NSCLC-Radiomics-Genomics (51 out of 89 patients) [22], QIN Lung CT (35 out of 47 patients) [23] [24], RIDER Lung CT (12 out of 32 patients) [25], RIDER Pilot (4 out of 8 patients) [26], SPIE-AAPM Lung CT Challenge (43 out of 77 patients) [27], and TCGA-LUAD (13 out of 69 patients) [28]. Demographic and clinical characteristics varied across datasets. For instance, Lung-Fused-CT-Pathology comprised 83.3% males with a mean age of 74.8 ± 10.3 years and a mean tumor size of 12.2 ± 1.9 mm. NSCLC-Radiomics included 72.3% males (mean age: 66.2 ± 10.3 years) with histologic subtypes of squamous cell carcinoma (50.2%), large cell (30.1%), adenocarcinoma (12.3%), and NOS/NA (7.4%). LungCT-Diagnosis had 59.6% of patients aged ≥65 years, with 53.2% males predominantly at advanced stages (III/IV). Moreover, we categorized patients based on overall survival into Class 1 (survival > 4 years) and Class 0 (survival < 4 years), representing rich prognosis and poor prognosis, respectively.

All CT scans were quality-checked and confirmed to be artifact-free prior to preprocessing. LCa lesions were manually segmented into 3D Slicer by three trained physicians and reviewed by an experienced thoracic radiologist. A total of 1,218 RFs were extracted using Laplacian of Gaussian (LoG, sigma:1-0-mm, 2-0-mm, 3-0-mm, 4-0-mm, 5-0-mm) and wavelet (LLH, LHL, LHH, HLL, HLH, HHL, HHH, LLL) filters under varying parameter settings to capture diverse spatial and textural characteristics. Features were normalized using min–max scaling. Among the 12 datasets, only NSCLC-Radiogenomics, LungCT-Diagnosis, and NSCLC-Radiomics included documented OS data, totaling 499 patients. The remaining datasets were used exclusively in the SSL framework. Both SL and SSL strategies were applied across multiple ML models to predict OS in NSCLC patients.

### 2.2. Classification Analysis

As illustrated in Figure 1, the proposed pipeline presents a comprehensive framework for developing robust machine learning models based on RFs extracted from lung cancer (LCa) CT images. It encompasses image preprocessing, dimensionality reduction through feature selection (FSA) and attribute extraction (AEA) algorithms, classifier benchmarking, and rigorous validation under both SL and SSL paradigms.

**i) Mask Delineation and ii) Expert Verification.** Chest CT examinations were initially reviewed using the standard lung window with multiplanar reconstructions (MPRs) to identify nodules exhibiting features suggestive of malignancy, including irregular shape, spiculated or lobulated margins, abnormal attenuation, or rapid size progression. Suspicious lesions were manually delineated on every axial slice using 3D Slicer (version 5.8) by two board-certified thoracic radiologists working in consensus. To ensure consistency and minimize annotation errors, the resulting segmentation masks were independently reviewed and confirmed by an additional clinical expert. This dual-review protocol was implemented to enhance inter-observer reliability and ensure the accuracy of lesion localization, which is critical for downstream RF extraction. Cases in which clear tumor boundary definition was compromised—due to pleural effusion, bulky lymphadenopathy, atelectasis, extensive fibrosis, or other obscuring pathology—were excluded from further analysis.

**iii) CT Intensity Normalization.** To account for variations in CT scanner settings and patient anatomy, each CT image undergoes intensity normalization using the min-max method. This transformation scales the voxel intensity values into a uniform range, typically [0, 1], which stabilizes the input for RF extraction. Normalization also ensures that intensity-dependent features are comparable across patients and imaging centers.

**iv) RF Extraction.** Once normalized, the lung CT images were processed using the open-source PyRadiomics package, which adheres to the Image Biomarker Standardization Initiative (IBSI) guidelines to ensure reproducibility and consistency. PyRadiomics enables the extraction of a comprehensive set of 119 RFs from each image, capturing diverse characteristics such as tumor heterogeneity, morphology, and tissue organization—critical properties for differentiating malignancy subtypes and predicting clinical outcomes. Specifically, the extracted features included 19 first-order statistics (FO), 16 three-dimensional shape features, 10 two-dimensional shape features, 23 gray level co-occurrence matrix (GLCM) features, 16 gray level size zone matrix (GLSZM) features, 16 gray level run length matrix (GLRLM) features, 5 neighboring gray tone difference matrix (NGTDM) features, and 14 gray level dependence matrix (GLDM) features. These standardized features were derived from the normalized CT volumes and are designed to capture both global and local image patterns relevant to tumor biology.

**v) Data Splitting and vi) Normalization Strategy.** Following feature extraction, the datasets were divided into a five-fold cross-validation set and separate external testing sets. The NSCLC-Radiomics dataset, which includes OS outcomes, served as the primary training cohort and was subjected to five-fold cross-validation. To evaluate model generalizability, two independent datasets with OS outcomes—NSCLC-Radiogenomics and LungCT-Diagnosis, each originating from distinct clinical centers—were designated for external testing. Datasets lacking outcome information were excluded from supervised training and instead incorporated into the SSL process. To ensure methodological rigor and avoid data leakage, normalization parameters (e.g., min and max) were derived exclusively from the training folds (four divisions) and subsequently applied to the validation, unlabeled datasets, and test sets during evaluation.

**vii) SL approaches:** In the SL framework, the labeled NSCLC-Radiomics dataset was partitioned into five folds. In each iteration, four folds were used for training, and the remaining fold was used for validation, ensuring that every fold served as the validation set once. This process was repeated five times to complete the cross-validation cycle. For external testing, the trained model from each fold was also evaluated on two independent labeled datasets—NSCLC-Radiogenomics and LungCT-Diagnosis—to assess generalizability across different centers and patient populations. The following performance metrics—Accuracy, Precision, Recall, F1-score, Receiver Operating Characteristic – Area Under the Curve (ROC-AUC), and Specificity [29]—were reported as both average values and standard deviations across the five cross-validation folds and external test evaluations. Model selection was based on the highest performance across all metrics during five-fold cross-validation and externally validated using independent test sets.

**viii) SSL approaches:** In the SSL framework, the labeled NSCLC-Radiomics dataset was split into five folds. In each iteration, a logistic regression model was trained using four labeled folds and then used to generate pseudo-labels for the unlabeled datasets (e.g., LCTSC, LIDC-IDRI, Lung-Fused-CT-Pathology, NSCLC-Radiomics-Genomics, QIN LUNG CT, RIDER, SPIE-AAPM, and TCGA). Importantly, the remaining labeled fold was excluded from the pseudo-labeling process to prevent bias and data leakage. After pseudo-labeling, the model was retrained on the combined labeled and pseudo-labeled data from the four folds, and evaluated on the held-out validation fold and two external test sets—NSCLC-Radiogenomics and LungCT-Diagnosis—to assess the added value of unlabeled data in improving model generalization.

**ix) Dimensionality Reduction via FSA and AEA.** To address the high dimensionality of RFs and reduce the risk of overfitting, the pipeline incorporates two parallel strategies: FSAs and AEAs [30]. A total of 56 dimensionality reduction techniques (27 FSAs and 29 AEAs) are evaluated for their ability to isolate the most informative and non-redundant features. The 27 FSAs encompass three major categories. Filter-based methods include the Chi-Square Test (CST), Correlation Coefficient (CC), Mutual Information (MI), and Information Gain Ratio, which score features independently of classifiers. Statistical tests such as ANOVA F-Test, ANOVA P-value selection, Chi2 P-value, and Variance Thresholding (VT) are used to assess feature discriminativeness. Wrapper-based methods include Recursive Feature Elimination (RFE), Univariate Feature Selection (UFS), Sequential Forward Selection (SFS), and Sequential Backward Selection (SBS), which iteratively evaluate subsets using model performance. Embedded methods such as Lasso, Elastic Net, Embedded Elastic Net, and Stability Selection perform selection during training. Ensemble-based approaches, including Feature Importance by RandF (FIRF), Extra Trees, and Permutation Importance (Perm-Imp), capture complex non-linear relationships. Additional statistical control methods like False Discovery Rate (FDR), Family-Wise Error (FWE), and multicollinearity handling techniques like Variance Inflation Factor (VIF) are also employed. Finally, dictionary-based strategies use Principal Component Analysis (PCA) or sparse loadings for stability and interpretability. The features selected by different FSAs are listed in Supplementary Files 1 and 2 (Sheet 3) for both the SL and SSL frameworks.

AEAs offer a complementary strategy by transforming the feature space into lower-dimensional subspaces. The 29 AEAs include linear techniques such as PCA, Truncated PCA, Sparse PCA (SPCA), and Kernel PCA (with RBF and polynomial kernels), which identify uncorrelated projections of maximum variance. Independent Component Analysis (ICA) and its variant FastICA extract statistically independent latent variables. Factor Analysis uncovers hidden structure behind observed features, while Non-negative Matrix Factorization (NMF) yields interpretable parts-based decompositions. SL techniques like Linear Discriminant Analysis (LDA) maximize class separation in the transformed space. Advanced manifold learning techniques capture nonlinear structures and include t-SNE, UMAP, Isomap, Locally Linear Embedding (LLE), Spectral Embedding, Multidimensional Scaling (MDS), and Diffusion Maps. These are especially useful for visualizing complex relationships in high-dimensional radiomics space. Deep learning methods, such as shallow and deep autoencoders, enable data-driven feature compression through reconstruction optimization. Other strategies include Feature Agglomeration for hierarchical grouping, Truncated SVD for matrix decomposition, and projection-based techniques like Gaussian Random Projection, Sparse Random Projection, and Feature Hashing, offering scalable compression.

**x) Classification Algorithms.** Each reduced feature set—whether selected by FSA or derived by AEA—is evaluated using a diverse set of 27 classifiers (CAs). These include linear models such as Tree-based classifiers encompass Decision Trees, Random Forest (RandF), Extra Trees, Gradient Boosting (GB), AdaBoost (AB), and HistGradient Boosting (HGB), each utilizing ensemble learning to reduce variance and improve generalization. Meta-ensemble strategies such as Stacking, Voting Classifiers (hard and soft), and Bagging further boosted predictive robustness by aggregating the strengths of multiple base learners. Support Vector Machines (SVM) were implemented with various kernels to handle linear and non-linear classification, while k-Nearest Neighbors (KNN) provided a distance-based, instance-level approach. Several Naive Bayes variants (Gaussian, Bernoulli, Complement) were tested for their probabilistic simplicity and computational efficiency. Neural network-based Multi-Layer Perceptron (MLP) facilitated modeling of complex patterns in non-linear spaces, whereas gradient-boosted frameworks such as Light GBM (LGBM) and XGBoost (XGB) provided high-performance learning through gradient optimization and feature importance modeling. Additional classifiers included LDA, Nearest Centroid, Decision Stump, Dummy Classifier (DC), Gaussian Process Classifier (GP), and Stochastic Gradient Descent Classifier (SGDC) for diverse modeling strategies. The classification algorithms were optimized using five-fold cross-validation and grid search. The optimal hyperparameters for each model in both SL and SSL frameworks are reported in Supplementary Files 1 and 2, Sheet 4.

**xi) Assess Sensitivity of Top-Performing Models to Data Size in SL and SSL.** To examine the impact and sensitivity of data volume on the performance of top-performing models in both SL and SSL approaches, we conducted three experimental scenarios. In the first scenario, based on the SL framework, we randomly selected 10% of the labeled NSCLC-Radiomics dataset for training and progressively increased the labeled portion in 10% increments until the full dataset (100%) was utilized. This allowed us to observe how model performance scaled with increasing availability of labeled data. In the second scenario, representing the SSL approach, we retained the entire labeled dataset and gradually introduced unlabeled data. Starting with 10% of the unlabeled pool, we incrementally added 10% more in each step until 100% of the unlabeled data was incorporated. This setup evaluated how additional unlabeled data contributes to model performance when the labeled set remains constant. In the third and final scenario, we simultaneously increased both labeled and unlabeled data. The experiment began with 10% of the labeled NSCLC-Radiomics data and 10% of the unlabeled data, increasing both in tandem by 10% steps until the full datasets were used. This setting assessed how balanced growth in both labeled and unlabeled data influences learning outcomes. Together, these experiments provided a comprehensive view of model robustness and adaptability to varying data volumes under both SL and SSL paradigms. In all three scenarios, only external testing metrics were reported, as the internal training and validation splits varied dynamically with each change in data volume. Reporting performance solely on fixed external test sets ensured consistency and fairness in comparing models across different data conditions.

**xii) Feature Importance Investigation by SHAP.** To investigate and interpret the contribution of individual RFs to classification outcomes, we applied SHapley Additive exPlanations (SHAP) to a set of 25 top-performing combinations of ML models and FSAs or AEAs. These combinations were selected based on their superior predictive performance across the accuracy metric. For each selected combination, SHAP values were computed to quantify the marginal contribution of each feature to the model's output, separately for class 0 (survival < 4 years) and class 1 (survival > 4 years) instances. To consolidate these insights, we then averaged the SHAP values for each feature across all combinations within each class. This aggregation allowed for a more stable and comprehensive understanding of feature importance patterns. Finally, we visualized the averaged SHAP values using heatmaps, enabling a comparative interpretation of feature contributions between the two classes. This approach not only enhances model transparency and interpretability but also reinforces the biological plausibility and diagnostic relevance of the selected features in LCa characterization.

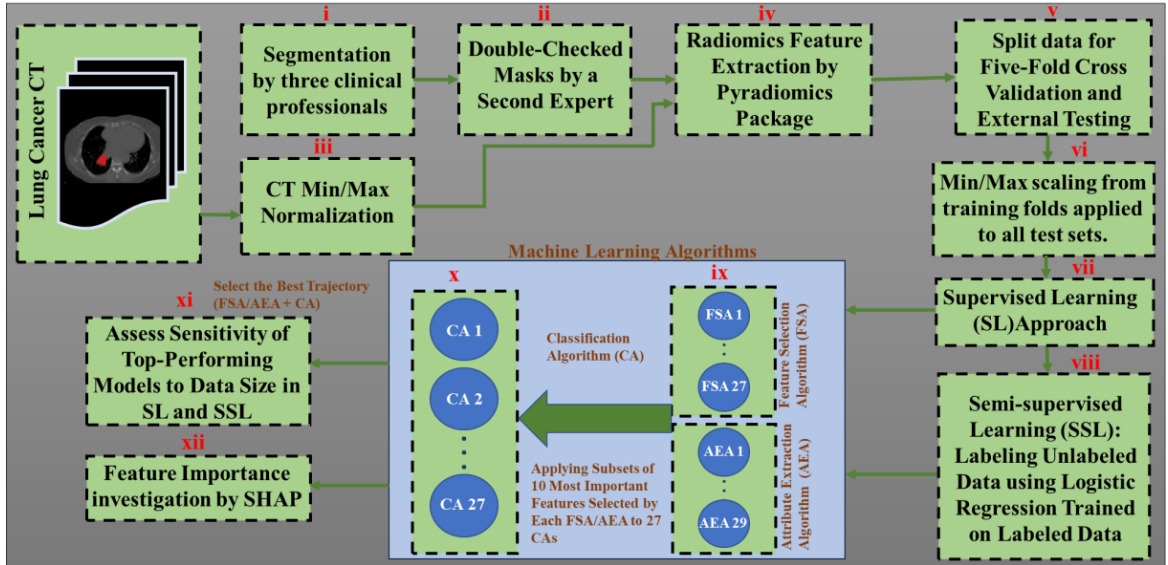

**Fig. 1.** Integrated Radiomics-Based Machine Learning Pipeline for Lung Cancer Classification and Prognosis. The pipeline integrates expert-guided tumor segmentation, min-max CT normalization, Image Biomarker Standardization Initiative (IBSI) (IBSI)-compliant radiomics feature extraction, dimensionality reduction, and classification algorithms. Supervised and semi-supervised learning strategies are implemented with five-fold cross-validation and external validation. Model robustness is evaluated through sensitivity analysis to data size, and feature importance is interpreted using SHAP.

## 3. Results

***SL vs. SSL.*** Figure 2 highlights the superior performance of select predictive models, with average accuracies from five-fold cross-validation presented for conciseness. Within the SSL framework, the FIRF + XGB combination demonstrated the highest OS prediction performance in LCa, achieving a cross-validation accuracy of $0.90 \pm 0.01$ and an external testing accuracy of $0.88 \pm 0.10$. In contrast, the top-performing model in the SL framework was RFE + LGBM, with a cross-validation accuracy of $0.78 \pm 0.02$ and an external testing accuracy of $0.87 \pm 0.01$. The difference in the cross-validation performance between the two frameworks was statistically significant ($p > 0.01$, paired t-test). These results suggest that incorporating unlabeled data through SSL promotes the learning of more robust and generalizable representations. Further analysis revealed a similar trend in AUC performance. In the SL setting, the best-performing model (RFE + LGBM) yielded a cross-validation AUC of $0.70 \pm 0.06$, but only $0.39 \pm 0.06$ on the external test set. In contrast, the SSL model (FIRF + XGB) delivered substantially higher values, with a cross-validation AUC of $0.87 \pm 0.02$ and an external test AUC of $0.73 \pm 0.02$.

Although Figure 2 highlights only a subset of top-performing classifiers and their associated FSAs/AEAs for brevity, Supplementary Files 1 and 2 provide a comprehensive overview of all 56 FSAs/AEAs combined with 27 classifiers, evaluated under both SL and SSL frameworks, respectively. These supplementary files include detailed performance metrics (Sheets 1 and 2)—Accuracy, Precision, Recall, F1-score, AUC, and Specificity—for both five-fold cross-validation and independent external testing, further substantiating the superior performance of SSL-based models. Moreover, Sheets 3 and 4 provide the optimal hyperparameters obtained through grid search optimization, as well as the selected features identified by the respective FSAs. These findings underscore the significant advantage of leveraging unlabeled data, since SSL methods consistently outperformed their SL counterparts across a broad range of classifier and FSA combinations. The observed performance differences between SSL and SL models were statistically significant, even after adjusting for multiple comparisons using the Benjamini-Hochberg procedure to control the FDR< 0.05.

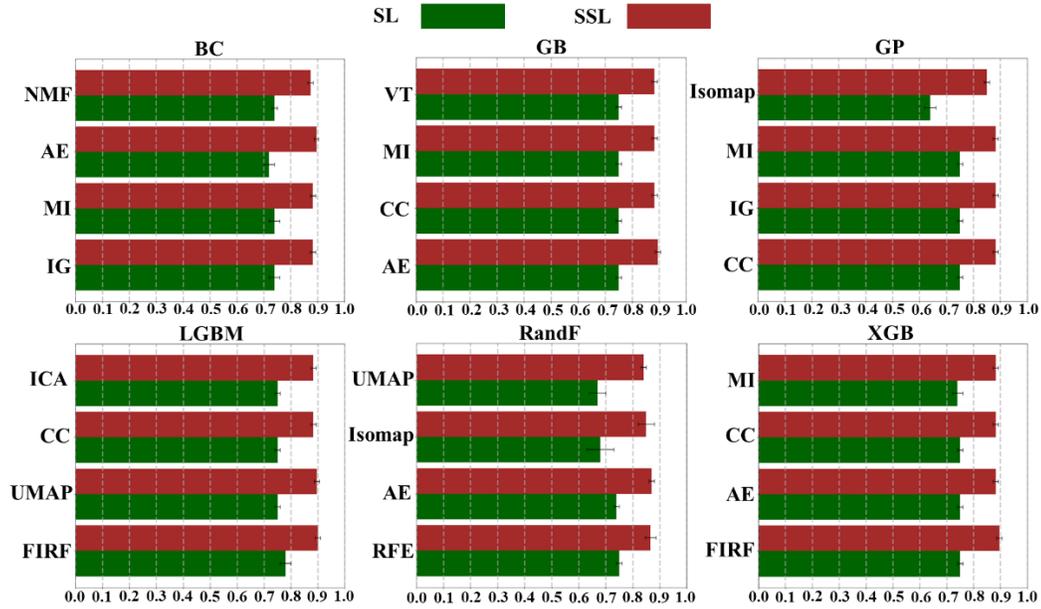

**Average Accuracy Across Five-Fold Cross-Validation**

**Fig. 2**. Bar plots illustrate the predictive performance of SL and SSL models for LCa survival prediction, with average 5-fold cross-validation accuracy shown for conciseness. Abbreviations: SL: Supervised learning, SSL: semi supervised learning, BC: Bagging classifier, GB: Gradient Boosting, GP: Gaussian Process Classifier, LGBM: Light Gradient Boosting Machine, RandF: Random Forest, XGB: XGBoost (eXtreme Gradient Boosting), NMF: Non-negative Matrix Factorization, AE: Auto encoder, MI: Mutual Information, IG: Information Gain, VT: Variance Thresholding, CC: Correlation Coefficient, ISOMAP: Isometric Mapping, RFE: Recursive Feature Elimination, FIRF: Feature Importance by RandF.

*Impact of Data Size on the Sensitivity of Top-Performing Models in SL and SSL.* As illustrated in Figure 3, we evaluated the sensitivity of top-performing models in both SL and SSL frameworks under varying training data sizes.

(i) In the first scenario, we assessed the best SL model using LGBM paired with RFE. The model was trained with progressively increasing proportions of labeled data—from 10% to 100%—while keeping the external test set fixed on NSCLC-Radiogenomics and LungCT-Diagnosis. Performance was averaged over 100 independent random splits for each step. Results revealed a consistent improvement in accuracy, starting at 0.74 with only 10% labeled data and reaching 0.87 with full labeled data. This confirms that model performance in SL is highly dependent on the quantity of labeled training data, particularly in the initial stages.

(i) The second scenario implemented the best SSL framework, using the same LGBM model but combined with the FIRF. Here, the entire labeled training set (NSCLC-Radiomics dataset) was supplemented with increasing proportions of unlabeled data. Remarkably, the model achieved an accuracy of 0.87 with just 10% unlabeled data, and performance remained stable as more unlabeled data was added, peaking slightly at 0.88 with 100% inclusion. This indicates that the SSL model was highly robust and benefited from unlabeled data without being overly sensitive to its proportion.

(iii) In the third scenario, both labeled and unlabeled data were jointly increased in 10% increments. Starting from 10% of total training data (a combined subset of labeled and unlabeled samples), the model began with an accuracy of 0.86 and quickly plateaued around 0.88 from the 60% mark onwards. This experiment, also conducted with LGBM + FIRF, suggests that a balanced and incremental addition of labeled and unlabeled data can rapidly approach the model's optimal performance. The plateau beyond 60% implies diminishing returns, indicating that a critical volume of data is sufficient for reliable learning.

In conclusion, the results clearly demonstrate that SSL strategies (Scenarios 2 and 3) offer significant advantages over purely SL approaches (Scenario 1) in settings with limited labeled data, as both SSL models achieved high and stable performance early on and exhibited reduced sensitivity to training data size.

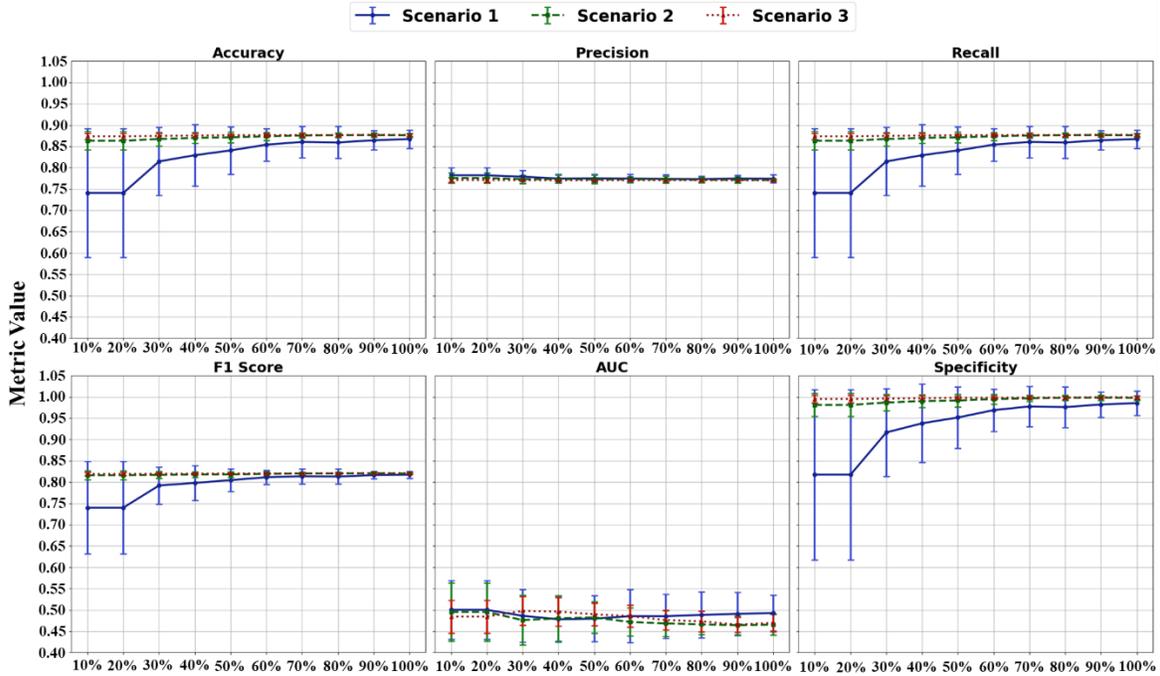

**Fig. 3**. Impact of training data size across three learning scenarios. Three scenarios were evaluated: Scenario 1 – supervised learning with increasing labeled data; Scenario 2 – semi-supervised learning with fixed labeled data and increasing unlabeled data; and Scenario 3 – SSL with jointly increasing labeled and unlabeled data. Performance trends demonstrate the advantage and stability of SSL approaches, especially when labeled data is limited.

***SHAP-Based Feature Importance in SL and SSL.*** In this analysis, model interpretability was explored using SHAP applied to the top 25 high-performing combinations of ML classifiers and FSAs/AEAs, identified based on five-fold cross-validation accuracy. These combinations involved five distinct classifiers paired with five different FSAs or AEAs.

In the SL setting, the heatmap in Figure 4(i) visualizes the average SHAP values for class 0 and class 1 across features selected by the top 25 model-FSA/AEA combinations in binary classification. Each row corresponds to a specific RF—such as textural, morphological, or wavelet-transformed features—while the two columns indicate the average absolute SHAP values for class 0 and class 1, respectively. The heatmap's color scale emphasizes relative importance: deep red highlights features contributing more strongly to class 0 predictions, while deep blue indicates higher influence on class 1. Notably, features such as Joint Entropy (GLCM_JEn, wavelet LHH), Kurtosis (FO_Ku, wavelet LHH), and Difference Entropy (GLCM_DiEn, wavelet LHH) showed strong contributions to class 1 predictions. Conversely, features like Gray Level Non-Uniformity (GLSZM_GLN, wavelet LLH), GLCM_JEn (wavelet LHH), and Small Dependence High Gray Level Emphasis (GLDM_SDHGLE, wavelet LLL) exhibited moderate influence on class 0. Overall, class 1 features tended to demonstrate stronger individual contributions, as reflected in more intense blue regions of the heatmap, aiding in both model interpretation and biomarker discovery.

In the SSL setting, Figure 4(ii) presents the corresponding average SHAP values for the same FSA process, applied to the top 25 model-FSA/AEA combinations. Each row again represents a selected RF, with SHAP values averaged for class 0 and class 1. Here, Gray Level Variance (GLSZM_GLV, LoG sigma:3.0) emerged as the most discriminative feature, strongly supporting class 1 predictions (SHAP: +0.500) while negatively impacting class 0 predictions (SHAP: −0.179). Additional key contributors to class 1 included Cluster Shade (GLCM_CS, wavelet LHH), GLCM_DiEn (wavelet LHH), and GLCM_JEn (wavelet LHH), all with substantial positive SHAP values. In contrast, High Gray Level Zone Emphasis (GLSZM_HGLZE, wavelet LLL) and GLCM_DiEn (wavelet LHH) showed negative SHAP contributions for class 0, suggesting they reduce model confidence for that class. These findings suggest that the SSL framework enhances the separability and consistency of key features—particularly texture and zone-based ones—by leveraging unlabeled data to refine decision boundaries, thereby improving both performance and interpretability.

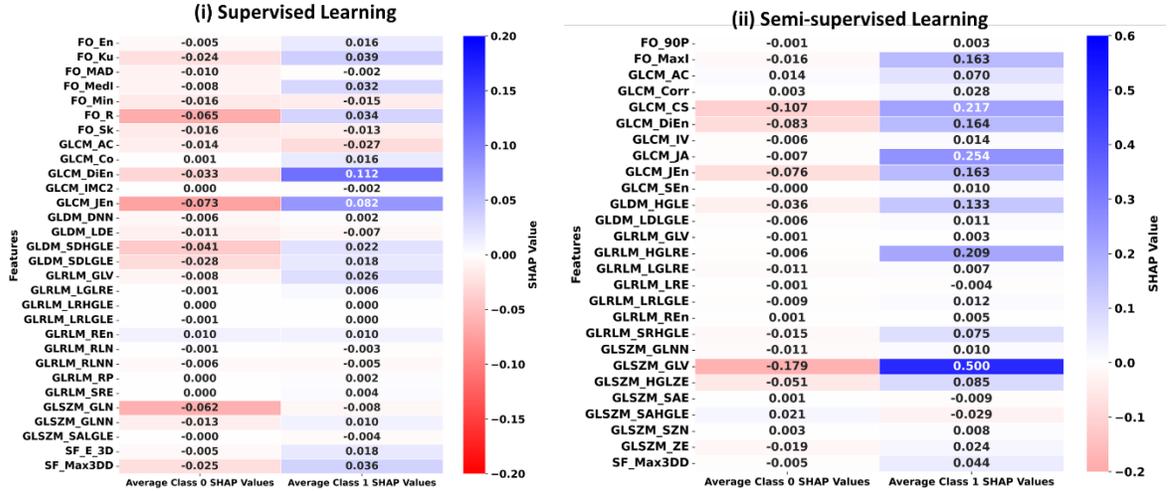

**Fig. 4.** Heatmaps of average SHAP values for feature importance in (i) supervised and (ii) semi-supervised learning settings. Red tones indicate higher contributions to class 0 (survival <4 years), while blue tones highlight features more influential for class 1 (survival >4 years). FO_90P: The 90th Percentile, FO_En: Entropy, FO_Ku: Kurtosis, FO_MAD: Mean Absolute Deviation, FO_MaxI: Maximum Intensity, FO_MedI: Median Intensity, FO_Min: Minimum, FO_R: Range, FO_Sk: Skewness, GLCM_AC: Autocorrelation, GLCM_Co: Contrast, GLCM_Corr: Correlation, GLCM_CS: Cluster Shade, GLCM_DiEn: Difference Entropy, GLCM_IMC2: Informational Measure of Correlation, GLCM_IV: Inverse Variance, GLCM_JA: Joint Average, GLCM_Jen: Joint Entropy, GLCM_Sen: Sum Entropy, GLDM_DNN: Dependence Non-Uniformity Normalized, GLDM_HGLE: High Gray Level Emphasis, GLDM_LDE: Large Dependence Emphasis, GLDM_LDLGLE: Large Dependence Low Gray Level Emphasis, GLDM_SDHGLE: Small Dependence High Gray Level Emphasis, GLDM_SDLGLE: Small Dependence Low Gray Level Emphasis, GLRLM_GLV: Gray Level Variance, GLRLM_HGLRE: High Gray Level Run Emphasis, GLRLM_LGLRE: Low Gray Level Run Emphasis, GLRLM_LRE: Long Run Emphasis, GLRLM_LRHGLE: Long Run High Gray Level Emphasis, GLRLM_LRLGLE: Long Run Low Gray Level Emphasis, GLRLM_Ren: Run Entropy, GLRLM_RLN: Run Length Non-Uniformity, GLRLM_RLNN: Run Length Non-Uniformity Normalized, GLRLM_RP: Run Percentage, GLRLM_SRE: Short Run Emphasis, GLRLM_SRHGLE: Short Run High Gray Level Emphasis, GLSZM_GLN: Gray Level Non-Uniformity, GLSZM_GLNN: Gray Level Non-Uniformity Normalized, GLSZM_GLV: Gray Level Variance, GLSZM_HGLZE: High Gray Level Zone Emphasis, GLSZM_SAE: Small Area Emphasis, GLSZM_SAHGLE: Small Area High Gray Level Emphasis, GLSZM_SALGLE: Small Area Low Gray Level Emphasis, GLSZM_SZN: Size-Zone Non-Uniformity, GLSZM_ZE: Zone Entropy, SF_E_3D: Elongation, SF_Max3DD: Maximum 3D Diameter

## 4. Discussion

This study presents a CT-only SSL framework for OS prediction in NSCLC, demonstrating superior performance, robustness, and clinical applicability compared to conventional SL approaches. Among all evaluated models, FIRF + XGB achieved the highest predictive accuracy and generalizability, supported by stable cross-validation performance and strong external test results.

The consistent superiority of SSL models can be attributed to their ability to utilize both labeled and unlabeled data, facilitating improved generalization and reducing overfitting, particularly in high-dimensional radiomics spaces. Unlike SL models that rely exclusively on limited annotated samples, SSL leverages pseudo-labeling to iteratively refine decision boundaries, enabling the model to learn a broader data distribution and better approximate underlying patterns. This mechanism explains the observed stability across different data sizes, where performance plateaued early—around 10–20% labeled data—and remained robust with increasing data volume.

Compared to a recent study [31] using SSL on LCa PET/CT RFs combined with clinical and deep learning features, our approach relies solely on standard-dose CT and multicenter data with external validation. That study showed improved performance when shifting from handcrafted RFs with SL to deep RFs with SSL, with PET-based deep RFs achieving 0.85 ± 0.05 accuracy and low-dose CT 0.83 ± 0.06. In contrast, our SSL model using only handcrafted RFs from standard dose CT achieved 0.90 ± 0.01 accuracy, demonstrating competitive performance without PET or deep features. While deep RFs may improve accuracy, they often lack interpretability [32]. Moreover, by leveraging standard-dose CT—an imaging modality that is widely available and routinely used in clinical practice—our approach enhances real-world applicability and facilitates broader clinical integration. Another study [33] showed that SSL consistently outperformed SL in PET/CT-based models for LCa survival prediction.

The enhanced generalizability of SSL models across multi-center datasets highlights their resilience to domain shifts caused by variations in acquisition protocols or institutional differences. Exposure to unlabeled data from diverse sources likely encouraged the learning of invariant features, enabling effective deployment across heterogeneous clinical environments.

SHAP-based interpretability analysis further reinforced SSL's effectiveness. Features emphasized in SSL—such as those related to gray-level entropy and wavelet-based texture—are known markers of tumor heterogeneity and have been associated with prognosis in NSCLC. Their prominence in SSL models suggests improved discrimination between classes due to enhanced exposure to broader patterns, which helps refine biologically relevant decision rules. Importantly, the SSL approach demonstrated reduced sensitivity to label scarcity, offering practical value in real-world settings where annotated datasets are limited or costly. The findings suggest that only a small fraction of labeled data is needed to achieve near-peak performance, making SSL highly suitable for integration into existing CT-based diagnostic workflows, especially in centers lacking access to PET or annotated imaging datasets.

The clinical significance of RFs in lung cancer CT imaging is multifaceted, with robust evidence supporting their roles in prognosis and treatment response prediction, particularly in survival outcomes. A comprehensive understanding of these CT RFs and their clinical implications is essential for adopting a precise and explainable approach. Features such as GLCM_JEn, GLCM_DiEn, and GLSZM_GLV are valuable in lung cancer CT imaging, with related clinical factors including tumor stage, histological type, and treatment received. These features aid in prognosis and survival prediction, with GLCM_JEn and GLCM_DiEn often associated with high-risk survival (class 0), and GLSZM_GLN linked to low-risk survival (class 1) in specific models. However, ongoing research is needed to address variability and validate their utility across diverse populations. Features like GLCM_JEn, GLCM_DiEn, and GLSZM_GL are important for distinguishing between benign and malignant lung nodules, with GLCM_JEn demonstrating moderate accuracy [34]. This non-invasive diagnostic capability can reduce unnecessary biopsies, enabling earlier detection and intervention, which is crucial for improving survival rates and increasing the likelihood of achieving class 1 outcomes. In the SL model, GLCM_JEn, GLCM_DiEn, and GLSZM_GLV strongly contribute to class 1 (low-risk survival), suggesting that higher values of these features are associated with prolonged survival. This aligns with existing literature indicating that higher entropy and kurtosis reflect tumor heterogeneity and intensity patterns, often indicative of more aggressive disease [35, 36]. Conversely, features like GLSZM_GLN and GLDM_SDHGLE, which contribute to class 0 (high-risk survival) in the SL model, suggest a worse prognosis, though this is unusual since GLSZM_GLN typically signifies heterogeneity and poorer outcomes [37]. In the SSL model, GLCM_JEn, GLCM_DiEn, and GLSZM_GLV strongly support class 1, reinforcing their association with prolonged survival, while GLRLM_LGLRE contributes to class 0, indicating less aggressive tumor characteristics.

Nonetheless, the pseudo-labeling strategy used in this study may be sensitive to early model errors, potentially leading to the propagation of incorrect labels. Incorporating confidence thresholds or consistency regularization in future work may mitigate this limitation. Moreover, while the current pipeline excludes clinical variables, integrating structured clinical data could enhance model robustness and provide more comprehensive prognostic insights.

From a clinical standpoint, the CT-only SSL model offers a low-cost, high-impact alternative for early risk stratification and personalized care planning. Its compatibility with standard imaging workflows, minimal dependence on expert annotation, and strong external performance support its potential for broad adoption. Future directions include prospective validation, incorporation of clinical covariates, and exploration of SSL frameworks in other cancer types to test generalizability. Moreover, future studies aim to develop radiological and biological dictionaries that bridge the interpretability gaps revealed by SHAP analysis, enabling alignment of RFs with established clinical scoring systems such as Lung-RADS (Lung Imaging Reporting and Data System), TNM staging, or WHO histologic classifications. Such integration would facilitate the translation of high-impact features into clinically meaningful descriptors, improving trust, explainability, and utility in real-world prognostic decision-making.

## 5. Conclusion

This study establishes SSL with pseudo-labeling as a robust, scalable strategy for OS prediction in NSCLC using only CT imaging. By effectively incorporating both labeled and unlabeled data, the framework overcomes limitations posed by scarce survival labels while maintaining high predictive performance. Compared to SL, SSL demonstrated superior accuracy, AUC, and resilience to variations in data size. SHAP-based analysis confirmed the enhanced interpretability and discriminative power of selected RFs. Clinically, the proposed approach supports early risk stratification and personalized treatment planning, offering a cost-effective, deployable solution for precision oncology in routine CT screening workflows.

**Data, Machine Learning Hyperparameters, and Code Availability.** All codes and tables are publicly shared at: *https://github.com/MohammadRSalmanpour/Robust-Semi-Supervised-CT-Radiomics-for-Lung-Cancer-Prognosis*

**Acknowledgment.** This study was supported by the Virtual Collaboration Group (VirCollab, www.vircollab.com) and the Technological Virtual Collaboration (TECVICO CORP.) based in Vancouver, Canada. We gratefully acknowledge funding from the Canadian Foundation for Innovation – John R. Evans Leaders Fund (CFI-JELF; Award

No. AWD-023869 CFI), as well as the Natural Sciences and Engineering Research Council of Canada (NSERC) Awards AWD-024385, RGPIN-2023-0357, and Discovery Horizons Grant DH-2025-00119.

**Conflict of Interest**. The co-authors Sonya Falahati, Mehrdad Oveisi, Mehdi Maghsudi, and Mohammad R. Salmanpour are affiliated with TECVICO CORP. The remaining co-authors declare no relevant conflicts of interest.

*Salmanpour et al*Personalized Prostate Cancer, Dictionary Version PM1. 0," *Journal of Imaging Informatics in Medicine,* pp. 1-22, 2025.

[33] M. Salmanpour, A. Gorji, A. Fathi Jouzdani, N. Sanati, R. Yuan and A. Rahmim, "Exploring Several Novel Strategies to Enhance Prediction of Lung Cancer Survival Time," in *2024 IEEE Nuclear Science Symposium (NSS), Medical Imaging Conference (MIC) and Room Temperature Semiconductor Detector Conference (RTSD)*, 2024.

[34] S.-H. Lee, H.-h. Cho, H. Y. Lee and H. Park, "Clinical impact of variability on CT radiomics and suggestions for suitable feature selection: a focus on lung cancer," *Cancer Imaging,* vol. 19, pp. 1-12, 2019.

[35] V. Nardone, P. Tini, P. Pastina, C. Botta, A. Reginelli, S. F. Carbone, R. Giannicola, G. Calabrese, C. Tebala and C. Guida, "Radiomics predicts survival of patients with advanced non-small cell lung cancer undergoing PD-1 blockade using Nivolumab," *Oncology Letters,* vol. 19, no. 2, pp. 1559-1566, 2020.

[36] M. R. Chetan and F. V. Gleeson, "Radiomics in predicting treatment response in non-small-cell lung cancer: current status, challenges and future perspectives," *European radiology,* vol. 31, pp. 1049-1058, 2021.

[37] H. J. Aerts, E. R. Velazquez, R. T. Leijenaar, C. Parmar, P. Grossmann, S. Carvalho, J. Bussink , R. Monshouwer, B. Haibe-Kains and D. Rietveld, "Decoding tumour phenotype by noninvasive imaging using a quantitative radiomics approach," *Nature communications,* vol. 5, no. 1, p. 4006, 2014.12